\documentclass[conference]{IEEEtran}
\pdfoutput=1
\IEEEoverridecommandlockouts

\usepackage[normalem]{ulem}
\usepackage{algorithm}
\usepackage {algpseudocode}
\algnewcommand{\LineComment}[1]{\State \(\triangleright\) #1}

\ifCLASSINFOpdf
   \usepackage[pdftex]{graphicx}
   \DeclareGraphicsExtensions{.pdf,.jpeg,.png}
\else
   \usepackage[dvips]{graphicx}
   \graphicspath{{.}}
\fi
\usepackage{tikz}
\usepackage{textcomp}
\usepackage{lipsum}

\newcommand\copyrighttext{%
  \footnotesize \textcopyright 2016 IEEE. Personal use of this material is
permitted.
  Permission from IEEE must be obtained for all other uses, in any current
or future
  media, including reprinting/republishing this material for advertising or
promotional
  purposes, creating new collective works, for resale or redistribution to
servers or
  lists, or reuse of any copyrighted component of this work in other works.}
\newcommand\copyrightnotice{%
\begin{tikzpicture}[remember picture,overlay]
\node[anchor=south,yshift=10pt] at (current page.south)
{\fbox{\parbox{\dimexpr\textwidth-\fboxsep-\fboxrule\relax}{\copyrighttext}}};
\end{tikzpicture}%
}

\usepackage{url}

\usepackage{lipsum}
\begin{document}
%
\title{PPCU: Proportional Per-packet Consistent Updates for Software Defined
Networks - A Technical Report}

\author{\IEEEauthorblockN{Radhika Sukapuram, Gautam Barua}
\IEEEauthorblockA{Indian Institute of Technology Guwahati\\
Guwahati, India 781039\\
Email: {\{r.sukapuram, gb\}@iitg.ernet.in}}}

\maketitle
\copyrightnotice

\begin{abstract}
In Software Defined Networks, where the network control plane can be
programmed by updating switch rules, consistently updating switches is a
challenging problem. In a per-packet consistent update (PPC), a packet
either matches the new rules added or the old rules to be deleted,
throughout the network, but not a combination of both. PPC must be preserved
during an update to prevent packet drops and loops, provide waypoint
invariance and to apply policies consistently. No algorithm exists today
that confines changes required during an update to only the affected
switches, yet preserves PPC and does not restrict applicable scenarios.  We
propose a general update algorithm called PPCU that preserves PPC, is
concurrent and provides an all-or-nothing semantics for an update,
irrespective of the execution speeds of switches and links, while confining
changes to only the affected switches and affected rules. We use data plane
time stamps to identify when the switches must move from the old rules to
the new rules.  For this, we use the powerful programming features provided
to the data plane by the emerging programmable switches, which also
guarantee line rate.  We prove the algorithm, identify its significant
parameters and analyze the parameters with respect to other algorithms in
the literature.
\end{abstract}


%
\section{Introduction}
\label{intro}
A network is composed of a data plane, which examines packet headers and
takes a forwarding decision by matching the forwarding tables in switches,
and a control plane, which builds the forwarding tables. A Software Defined
Network (SDN) simplifies network management by separating the data and
control planes and providing a programmatic interface to the control plane.

SDN applications running on one or more controllers can program the network
control plane dynamically by updating the rules of the forwarding tables
of switches to alter the behaviour of the network.  Every Rules Update
consists of a set of updates in a subset $S$ of the switches $B$ in a
network. Every update in switch $s_i \in S$  consists of two sets of rules
$R_1$ and $R_0$.  Rule set $R_0$ is to be deleted and rule set $R_1$ is to
be inserted. One of $R_0$ or $R_1$ may be NULL. The rules are denoted by
$s_i(R_1,R_0)$.  Every switch $s_i$,  $s_i \in S$, is called an
\textit{affected switch}. All other switches $(B-S)$ are unaffected
switches. The set of rules that is neither inserted nor deleted is an
unaffected set of rules and is denoted by $R_u$.

\begin{figure}[t]
 \caption{Need for PPC updates}
 \centering 
 \includegraphics[scale=0.4, trim=0 290 100 0]{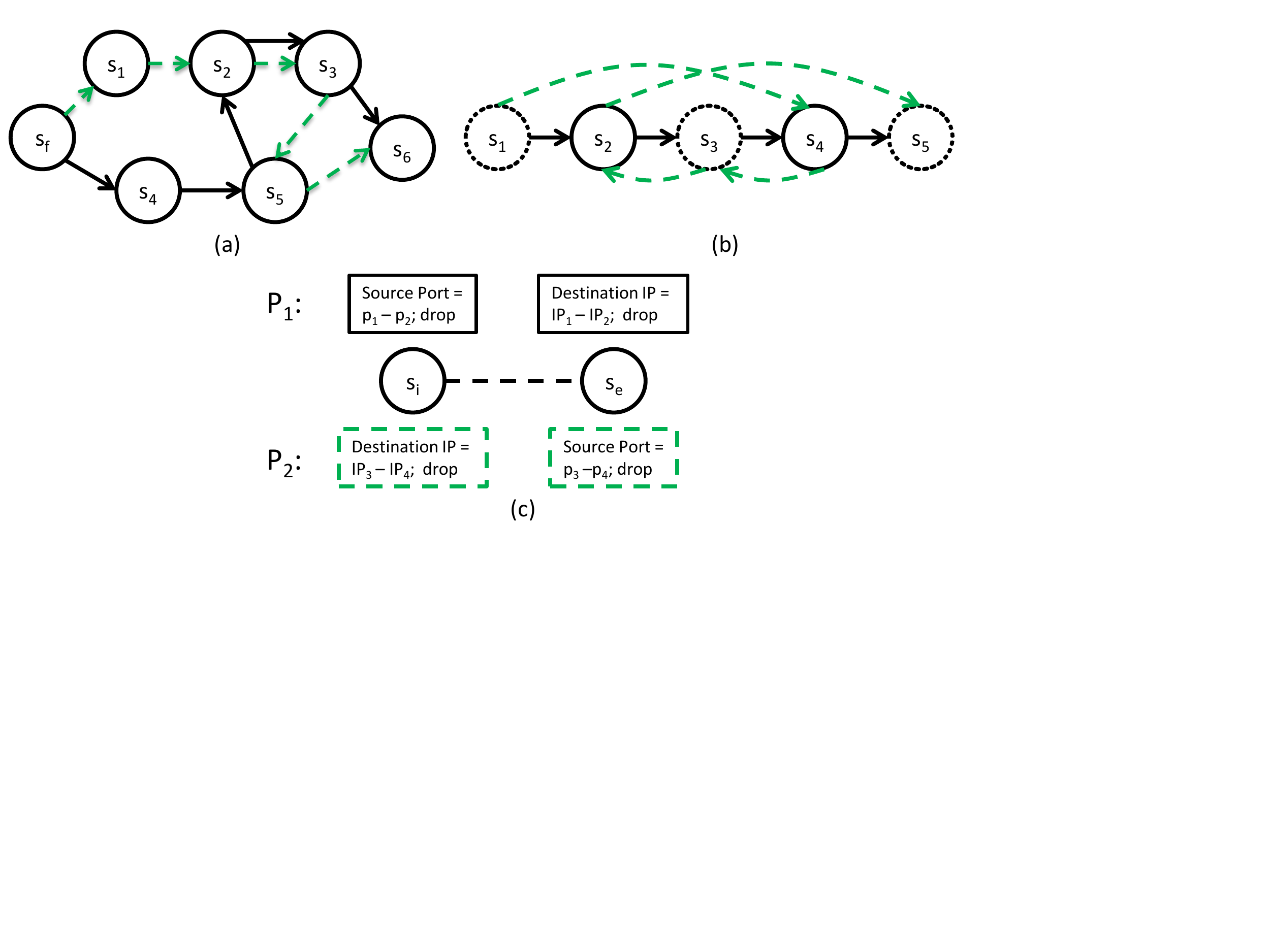} 
 \label{PPC} 
\end{figure}

Packets that traverse the network during the course of an update must not be
dropped (\textit{drop-freedom}) and must not loop (\textit{loop-freedom}) on
account of the update.
In Figure \ref{PPC}(a),  when switches are updated to change a route from
the old path (solid lines), to the new path (dotted lines), if $s_f$ is
updated first, packets arriving at $s_1$ will get dropped, while if $s_3$ is
updated first, packets arriving at $s_3$ will loop. Flows in a network may
need to pass through a number of waypoints in a certain order at all times
(\textit{waypoint invariance}).  In Figure \ref{PPC}(b), $s_1$, $s_3$ and
$s_5$ are the waypoints to be traversed in that order and this cannot be
achieved by sequencing the updates in any order\cite{189032}. While the
above three properties can be preserved in many update scenarios by finding
a suitable order of updates \cite{189032, ludwig2014good, dsn16, mattosreverse}, the example in Figure \ref{PPC}(c) cannot be
solved by suitably sequencing updates, in any scenario. Initially the
network has a policy $P_1$ which requires an ingress switch $s_i$ to drop
packets whose ports are in the range $p_1-p_2$ and $s_e$, an egress switch,
to drop packets whose destination addresses are in the range $IP_1$ to
$IP_2$. Policies may be distributed over various switches in an
SDN to reduce load on middleboxes or on links to them \cite{ben2015enforsdn}
or due to lack of rule space on a switch \cite{nguyen2016rules}.
When the
policy is changed to $P_2$, the network administrator desires either $P_1$
or $P_2$ to be applied to any packet $p$ and not, for example, $P_1$ at
$s_i$ and $P_2$ at $s_e$. We call this property \textit{pure per-packet
consistency (pure PPC)}.

In order to preserve all the properties for all scenarios, every Rules
Update must be per-packet consistent
\cite{DBLP:conf/sigcomm/ReitblattFRSW12} (PPC): every packet $P$ travelling
through the network must use either rule $r_0 \in R_0$ in every affected
switch $s_i \in S$  or rule $r_1 \in R_1$ in every affected switch $s_i$,
and never a combination of both, during a Rules Update. Other rules in the
switches are to be used if they match.  The number of packets that get
subjected to any of the inconsistencies can be quite large because the time
to update a few rules at a switch, which is of the order of milliseconds
\cite{DBLP:conf/sigcomm/JinLGKMZRW14}, is far greater than the time to
switch a packet out of a switch, which is of the order of nanoseconds, given
that the line rate of switches is 10-100 Gbits/s on 10-100 ports
\cite{packettx} and assuming an average packet size of 850 bytes
\cite{benson2010understanding}.

The basic algorithm that preserves PPC is a \textit{two-phase update} (2PU)
\cite{DBLP:conf/sigcomm/ReitblattFRSW12}: All the rules in all the switches
check for version numbers. All the incoming packets are labelled with the
appropriate version number by all the ingresses. To update a set of rules
from $v_0$ to $v_1$, first the new $v_1$ rules and a copy of
the unaffected rules are installed to check for $v_1$, in every internal (
non-ingress) switch. Now all the ingresses start labelling all the packets
as $v_1$. All $v_1$ packets match only the new rules and the $v_0$ packets
in the network match only the old rules, preserving PPC. After all the $v_0$
packets exit the network, the controller deletes the old rules on all the
switches. 2PU affects all the switches in the network.

\textit{We propose an update algorithm called Proportional Per-packet
Consistent Updates, PPCU, that is general, preserves PPC,
confines changes to only the affected switches and rules and provides an
all-or-nothing semantics for a Rules Update, regardless of the execution
speeds of switches and links}.

The summary of PPCU is as follows: Let the latest time at which
all the switches in $S$ install the new rules be $T_{last}$.  Each affected
switch examines the time stamp, set to the current time by the ingresses, in
each data packet. If its value is less than $T_{last}$, it is switched
according to the old rules while if it is greater than or equal to
$T_{last}$, it is switched according to the new rules. 
\section{Motivation and related work}
\label{moti}
\textbf{Why preserving PPC is important:}
Proper installation of policies require preservation of PPC and no less.
Middle box deployment in SDN \cite{ben2015enforsdn} \cite{gember2015opennf}
rely on packets traversing middle boxes in a certain order \cite{joseph2008policy}
\cite{fayazbakhsh2013flowtags}.  Languages \cite{arashloo2015snap} that
enable SDN application developers to specify and use states, also require
the order of traversal of switches that store those states to be preserved
during an update. These examples emphasize the importance of preserving
waypoint invariance for \textit{all} scenarios and therefore of PPC.  For
many algorithms that preserve more relaxed properties, a PPC preserving
update is a fall-back option, for example, if they cannot find an order of
updates that preserve waypoint invariance \cite{189032} or if their Integer
Linear Programming algorithms do not converge \cite{flip}.

\textbf{Why the number of switches involved in the update must be less:}
Algorithms that perform a Rules Update while preserving PPC either 1) need to
update all the switches $B$ in the network
\cite{DBLP:conf/sigcomm/ReitblattFRSW12} \cite{sukapuram2015enhanced} or 2)
identify paths affected by $S$ and update \textit{all} the switches along
those paths \cite{189032} \cite{DBLP:conf/sigcomm/KattaRW13} or 3) update
only $S$ and all the \textit{ingresses} \cite{CCU}
\cite{luo2015consistency}, regardless of the number of elements in $S$ and
regardless of the number of rules to be updated. 

If paths affected by a Rules Update are identified (category 2), every
switch in the path, which may be across the network, needs to be changed,
even to modify one rule on one switch. If one rule affects a large number of
paths, such as ``if TCP port=80, forward to port 1'', computing the paths
affected is time consuming \cite{189032}. In update methods of categories 1
and 2, the number of rules used in \textit{every} switch modified during the
update doubles, as the new and old rules co-exist for some time. Hence the
update time is disproportionately large, even if the number of rules updated is small.

If rules can be proactively installed on switches to prevent a control
message being sent to a central controller for every missing rule
\cite{yeganeh2013scalability}, one controller can manage an entire data
centre.  In a data centre with a Fat Tree topology with a k-ary tree
\cite{Al-Fares:2008:SCD:1402946.1402967} where the number of ports $k=48$,
the number of ingress switches is about $92\%$ of the total number of
switches. Even to modify one rule in a switch, all of the ingresses will
need to be modified, for algorithms of category 3.  

Updating only switches in $S$ instead of $S$ and other switches in $B-S$
will reduce the number of controller-switch messages and the number of
individual switch updates that need to be successful for the Rules Update to
be successful.  We wish to characterize this need as \textit{footprint
proportionality (FP)}, which is the ratio of the number of affected switches
for a Rules Update to the number of switches actually modified for the
update.  In the best case the FP is 1, and there is no general update
algorithm that achieves this while maintaining PPC, as far as we know. 

\textbf{Why using rules with wildcards is important:} 
Installing wildcarded rules for selected flows or rule compression in TCAM
for flows or policies \cite{zhang2014adaptable} reduces both space occupancy
in TCAM \cite{luo2014fast} and prevents sending a message to the controller
for every new flow in the network \cite{yeganeh2013scalability}.
\cite{iyer2013switchreduce} provides solutions for the tradeoff between
visibility of rules and space occupancy.  Thus in a real network, using
wildcarded rules is inevitable. Many of the existing solutions do not
address rules that cater to more than one flow
\cite{DBLP:conf/sigcomm/JinLGKMZRW14} \cite{189032}. A Rules Update may
require only removing rules and not adding any rule or vice versa, on one or
more switches in a network.  Though \cite{luo2015consistency} addresses
this, it needs changes to all the ingresses for all the rules for any update and limits
concurrency.  We wish to have $FP = 1$ for all the update scenarios
possible.

\textbf{Why concurrent updates must be supported:}
Support of multiple tenants on data center networks require updating
switches for virtual to physical network mapping
\cite{banikazemi2013meridian,al2014openvirtex}. Applications that control
the network \cite{benson2011microte}, \cite{al2010hedera},
\cite{heller2010elastictree} and VM and virtual network
migration\cite{ghorbani2014transparent} will require frequent updates to the
network, making large concurrent updates common to SDNs.  Existing
algorithms that preserve PPC and allow concurrent updates limit the
maximum number of concurrent updates \cite{CCU} \cite{luo2015consistency}
\cite{189032}, with the last one becoming impractical if the number of paths
affected by the update is large. 2PU
\cite{DBLP:conf/sigcomm/ReitblattFRSW12} does not support concurrency.

\textbf{Why algorithms that use data plane time stamping are inadequate:}
Update algorithms that use data plane time stamping either cater to only
certain update scenarios or require very accurate and synchronous time
stamping \cite{mizrahi2016software} or do not guarantee an all-or-nothing
semantics for Rules Updates as they depend on an estimate for the time at
which the update must take place \cite{timed} \cite{OpenflowSpec}, thus
relying on the execution speeds on switches and links to be predictable.

\textbf{Our contributions in this paper are: }\\
1)\textbf{ We specify and prove an update algorithm PPCU that:} a) requires modifications
to only the affected switches $S$ and affected rules, while preserving PPC.
b) allows practically unlimited concurrent non-conflicting updates.
c) allows all update scenarios:
this includes Rules Updates that involve only deletion of rules or only insertion
of rules or a combination of both and Rules Updates that involve forwarding
rules that match more than one flow.
d) expects the switches to be synchronized to a global
clock but tolerates inaccuracies.
e) updates all of $S$ or none at
all, irrespective of the execution speeds of the switches or links.\\
2) We prove the algorithm, analyze its significant parameters and find
them to be better than comparable algorithms.\\
3) We illustrate that the algorithm can be implemented at line
rate.
\section{Using data plane programmability:}
\label{DP}
With the advent of line rate switches whose parsers, match fields and
actions can be programmed in the field using languages such as P4
\cite{bosshart2014p4} and Domino \cite{packettx}, it has become possible to
write algorithms for the data plane in software.  Programmable switch
architectures such as Banzai \cite{packettx}, RMT\cite{bosshart2013forwarding},
FlexPipe\cite{ozdag2012intel} (used in Intel FM6000 chip set\cite{fm6000}) and
XPliant\cite{xpliant} aid this and languages to develop SDN applications
leveraging these abilities
\cite{schlesinger2014concurrent}\cite{arashloo2015snap} and P4 applications
\cite{kattahula} \cite{sivaraman2015dc} indicate acceptability.

Data plane programmability is important to PPCU for several reasons: 1) It
is easier to change protocols to add and remove headers to and from packets
2) It is possible to process packet headers in the data plane, still
maintaining line rate 3) Once the packet processing algorithm is specified
in a language, the language compiler finds the best arrangement of rules on
the physical switch \cite{jose2015compiling}. It is possible that rules that
require a ternary match may be stored in TCAM and exact match rules in SRAM
\cite{bosshart2013forwarding} or the former in a Frame Forwarding Unit and
the latter in a hash table \cite{ozdag2012intel}. 4) The actual match-field
values in the rules, the actions, the parameters associated with actions and
the variables associated with rules can be populated and modified at run time.
How to do this is a part of the control plane and is proposed to be
``Openflow 2.0'' \cite{bosshart2014p4}. Specifications such as
SAI\cite{SAISpec} or Thrift \cite{Thrift} or the run-time API generated by
the P4 compiler \cite{P4Spec} \cite{sivaraman2015dc} may be used at present.
We assume that switches support data plane programmability and can be
programmed in P4. 

\section{Model Description}
\label{model}
\textbf{Description of a programmable switch model:}
The abstract forwarding model advocated by protocol independent programmable
switches \cite{P4Spec} consists of a parser that parses the packet headers,
sends them to a pipeline of ingress match-action tables (the
\textit{ingress pipeline}) that in turn consist of a set of match-fields
and associated actions, then a queue or a buffer, followed by a pipeline of
egress match-action tables (the \textit{egress pipeline}). In
this section, we describe only the features of P4 that are relevant to PPCU.

A P4 program consists of definitions of 1) packet header fields 2) parser
functions for the packet headers 3) a series of \textit{match-action tables}
4) \textit{compound actions}, made of a series of \textit{primitive} actions
and 5) a control flow, which imperatively specifies the order in which
tables must be applied to a packet.  Each match-action table specifies the
input fields to match against; the input fields may contain packet headers
and \textit{metadata}. The match-action table also contains the actions to
apply, which may use metadata and \textit{registers}. Metadata is memory
that is specific to each packet, which may be set by the switch on its own
(example: value of ingress port) or by the actions. It may be used in the
match field to match packets or it may be read in the actions. Upon entering
the switch for the first time, the metadata associated with a packet is
initialised to $0$ by default. A register is a stateful resource and it may
be associated with each entry in a table (not with a packet).  A register
may be written to and read in actions. 

P4 provides a set of primitive actions such as $modify\_field$ and
$add\_header$ and allows passing \textit{parameters} to these actions,
that may be metadata, packet headers, registers etc.  When the primitive
action $resubmit$ is applied to the ingress pipeline, a packet completes its
ingress pipeline and then resubmits the \textit{original} packet header and
the possibly modified metadata associated with the packet, to the parser;
the metadata is available for matching. If there are multiple $resubmit$
actions, the metadata associated with each of them must be made available to
the parser when the packet is resubmitted.  Similarly, when $recirculate$ is
applied to a packet in the egress pipeline, the packet, with its header
modifications and metadata modifications, if any, is posted to the parser.
Conditional operators are available for use in compound actions
to process expressions (we use if statements in the paper to improve
readability).

While this specifies the \textit{definition} of the programmable regions of
the switch, actual rules (table entries) and the parameters to be passed to
actions need to be specified and is described below. These need to be
populated by an entity external to this model - the controller, through the
switch CPU. This is facilitated by a run-time API.

\textbf{Description of Rules and packets:}
A switch has an ordered set of rules $K$, consisting of $n$ rules $r_1, r_2,
...r_n$.  Each rule $r$ has three parts: a priority $P$, a match field $M$
and an action field $A$ and is represented as $r = [P, M, A]$. Let $r_1 =
[P_1, M_1, A_1]$ and $r_2 = [P_2, M_2, A_2]$. $r_1 \prec r_2$ if $P_1 >
P_2$. $M$ must be as per the match field defined in the table
and $A$ must be one of the actions associated with the table.  A packet $P$
only has match fields. An incoming packet is matched with the match fields
of the rules in the first table specified in the control flow of the switch
and the action associated with the highest priority rule that matches it is
executed. The packet is then forwarded to the next match-action table, as
specified in the control flow.

\begin{figure}[t]
 \caption{Dependencies among rules}
 \centering 
 \includegraphics[scale=0.4, trim=0 300 100 0]{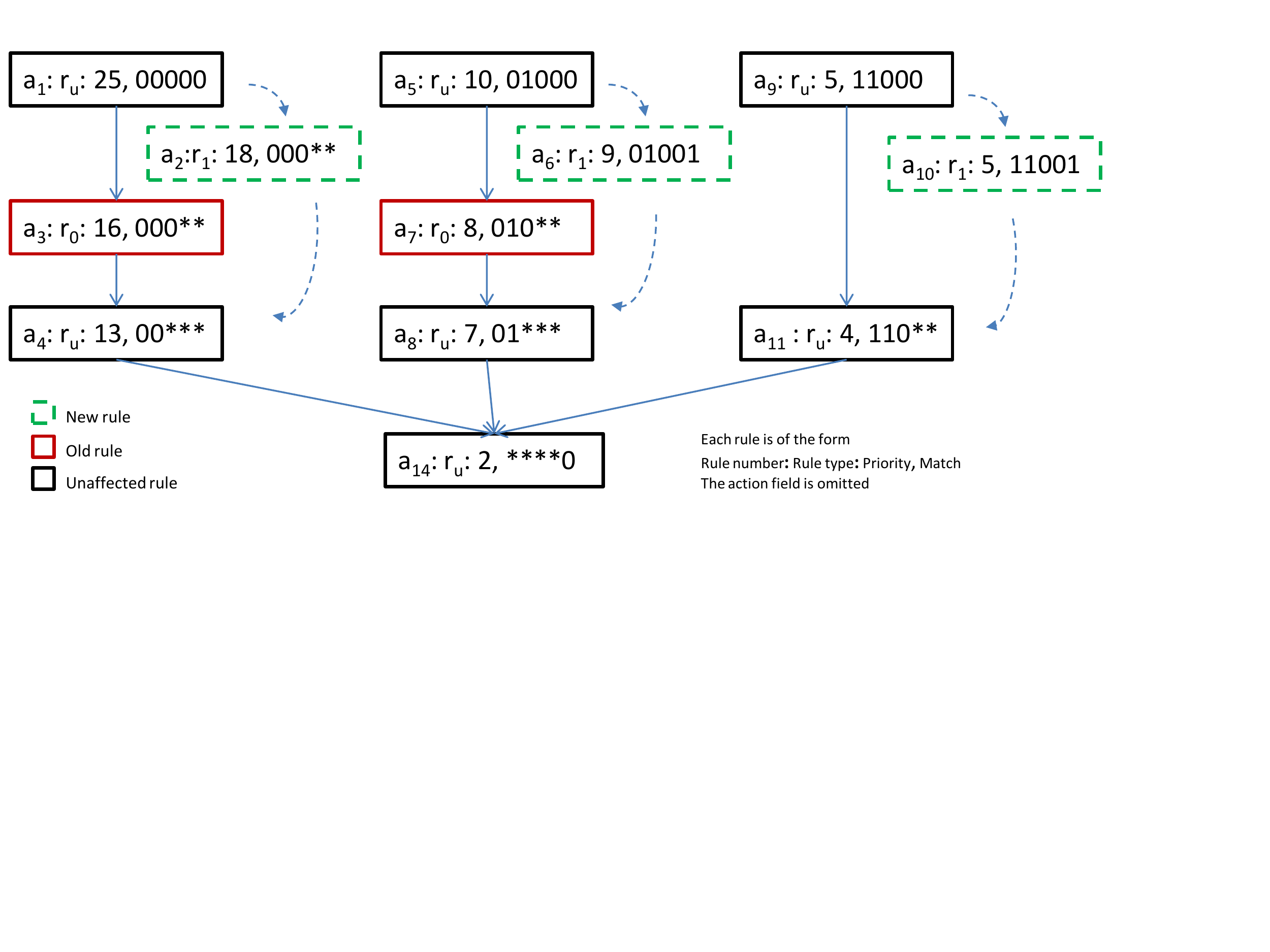} 
 \label{order} 
\end{figure}

In $K$, some rules may be dependent on the other. For example, in Figure
\ref{order}, the rules $a_1$, $a_3$, $a_4$ and $a_{14}$ depend on each
other. The solid arrows show the existing dependencies among rules and the
dotted arrows, the dependencies after a Rules Update.

$r$ with suitable subscripts or superscripts denotes an individual rule,
while $R$ with suitable subscripts or superscripts denotes a set of rules,
with $r_0 \in R_0$, $r_1 \in R_1$ and $r_u \in R_u$, throughout the paper.
A packet that matches any $r \in R_1$ in any $s \in S$ is called a
\textit{new} packet. A packet that matches any $r \in R_0$ in any $s \in S$
is called an \textit{old} packet and all old and new packets are called
\textit{affected packets}. The \textit{first} affected switch containing at
least one of $R_1$ or $R_0$ that a packet matches is denoted as $s_f$. 

\textbf{Disjoint Rules Updates:}
\label{disjoint}
Let $P_1$ be the set of all packets that match at least one of $R_0$ or
$R_1$ of a Rules Update $U_1$ and let $P_2$ be the set of all packets that
match at least one of $R_0$ or $R_1$ of another Rules Update $U_2$. $U_1$
and $U_2$ are said to be disjoint or non-conflicting if $P_1$ and $P_2$ are disjoint. If $U_1$
and $U_2$ are disjoint, they can occur concurrently and every packet in the
network will be affected by at most one of the updates $U_1$ or $U_2$.  For
example, let $R_1 = \{[5, 1000, forward\ 10]\}$. Let $s_i(null,R_1)$ be the
only update in $U_1$.  Let $R_0 = \{[5, 1***, forward\ 5]\}$. Let
$s_j(R_0,null)$ be the only update of $U_2$.  $U_1$ and $U_2$ conflict. Let
$R_1' = \{[5, 1111, forward\ 10]\}$. Let $s_i(null,R_1')$ be the only
update in $U_3$.  $U_1$ and $U_3$ are disjoint. One Rules Update can consist
of one or more disjoint updates. In Figure \ref{order}, $a_3$ and $a_7$ need
to be deleted and $a_2$, $a_6$ and $a_{10}$ need to be installed. They can
all be part of the same Rules Update.

\textbf{Relationship between the old and the new rules:}
\label{new-cases}
Two ordered sets of rules $A$ and $B$ are said to be \textit{match-field
equivalent} if the total set of packets that they are capable of matching
are identical.  Thus $A = \{[P_1, 0000,A_1], [P_2, 01**, A_2]\}$ and $B =
\{[P_1, 0000,B_1], [P_2, 0100, B_2], [P_3, 01**, B_3]\}$ are match-field
equivalent. 

The set of rules in $A$ for which no match-field equivalent rules of
priority equal to or lower than $A$ exists in $B$ is called the special
difference of $A$ and $B$, denoted $A - B$.  If $R_1$ = $\{[P_1, 0000, A_1],
[P_2, 00**, A_2]\}$ and $R_0$=$\{[P_2, 000*, B_1]\}$, where $P_1 > P_2$,
$R_1 - R_0 = {[P_2, 00**, A_2]}$.  The set of rules in $A$ for which no
match-field equivalent rules of priority \textit{equal to or greater than}
$A$ exists in $B$ is called the inverse special difference of $A$ and $B$,
denoted $A \sim B$.  If $R_0$ = $\{[P_2, 00**, A_2]\}$ and $R_1 = \{[P_1,
000*, B_1]\}$, where $P_1 > P_2$, $R_0 \sim R_1 = {[P_2, 00**, A_2]}$.

Either or both of $R_0 \sim R_1$ and $R_1 - R_0$ may not be $\phi$ for every
affected switch.  For example, in Figure \ref{order}, $\{a_7\} \sim \{a_6\}
\neq \phi$ and $a_{10}$ has no corresponding old rule. Thus, both $R_0 \sim
R_1$ and $R_1 - R_0$ are not $\phi$.  This has implications for the update
algorithm, as discussed in section \ref{main-challenges}.

\section{Challenges in a Rules Update} 
\label{main-challenges}
Providing an FP of 1 is difficult because multiple disjoint updates may be
combined into one Rules Update and a switch rule may belong to more than one
flow, due to usage of wild carded rules. Thus for packets belonging to
different flows the first affected switch $s_f$ may be different - that is,
there may be more than one $s_f$ belonging to a Rules Update. Neither the
controller nor the switches know the paths that are affected by the update
or the $s_f$ for each path.  Due to these reasons, firstly, the controller
cannot instruct a specific switch to relabel packets to switch to the new
version of rules, as is done in some of the update algorithms \cite{189032}.
Secondly, all switches must have the same algorithm, regardless of their
position in a flow. Moreover, the rules to be deleted and installed and the
unaffected rules depend on each other in complex ways, as discussed in
\ref{new-cases}, causing difficulty in maintaining PPC. \textit{The key
insight is that all the affected switches must know if the rest of the
affected switches and itself are ready to move to the next version, and
after that, if a packet crosses \textbf{its} $s_f$, it must store that fact
in its header ; the switch examines $TS$, the time stamped at the ingress on
every packet, to understand the readiness and stores the result in two
one-bit fields $f_{p1}$ and $f_{p2}$ in the packet.}

Since the controller-switch network is asynchronous, when a controller sends
messages to switches, all switches will not receive them at the same time,
or complete processing them and respond to the controller at the same time
or they may not respond at all.  \textit{Solving race conditions arising out
of these scenarios is possible if the values of $f_{p1}$ and $f_{p2}$ which
are set by $s_f$ determine whether they match $R_0$ or $R_1$ or $R_u$,
regardless of what the states of the switches in its path are}. Also, the
clocks at the switches may be out of sync to the extent the time
synchronising protocol would permit. In the next section, we shall state the
algorithm first and then describe the measures taken in them to address the
above challenges.
\section{Algorithm for concurrent consistent updates}
\textbf{Data plane changes at the ingress and egress switches:}
Each packet $p$ entering the network has \textit{a time stamp field and two
one bit fields $f_{p1}$ and $f_{p2}$} added to it and removed from it
programmatically, at the ingress and the egress, respectively. All ingresses
set $TS$ to the current time at the switch and $f_{p1}$ and $f_{p2}$ to $0$
for all the packets entering it \textit{from outside the network}.  Altering
the value of $TS$, wherever required in the update algorithm, occurs only
after it is thus set. 

\textbf{Notations used:}
$flag$ indicates if the rule is new ($NEW$), old ($OLD$) or
unaffected ($U$).  $f_{p1}$, when set to $1$, indicates that the packet must
be switched only according to $OLD$ rules, where they exist and $f_{p2}$,
when set to $1$ indicates that the packet must be switched only according to
$NEW$ rules, where they exist.  The metadata bit $f_1$ when set to $1$ for a
packet indicates that this packet must not be matched by a $NEW$ rule.  The
metadata bit $f_2$ when set to $1$ for a packet indicates that this packet
must not be matched by an $OLD$ rule.

\label{final_alg}
\begin{algorithm}
\caption{Algorithm at the controller}
\label{Controller}
\begin{algorithmic}[1]
\Procedure{Controller}{\null} 
\While {1}
  \If {$event$ = $app\_message\_received$}
    \For {each switch $s$ in $S$}
      \State \Call {send\_commit}{$v, R_0, R_1$} \label{commit_sent}
    \EndFor
  \ElsIf {$event$ = $R2C\_recd$} 
      \State {$T_l$ = time received in Ready To Commit}
      \If {R2C is received from all affected switches}
        \State {$T_{last}$ = the largest value of $T_l$ received } \label{tlm}
        \State {send CommitOK($v,T_{last}$) to all switches in $S$}
\label{commit_ok_sent}
      \EndIf
  \ElsIf {$event$ = $Ack\_COK\_recd$}
      \State {$T_a$ = time received in Ack Commit OK}
      \If {Ack COK is received from all affected switches}
        \State {$T_{del}$ = the largest value of $T_a$ received}
        \State {send\_DiscardOld ($v,T_{del}$) to all
affected switches}
      \EndIf
  \ElsIf {$event$ = $Discard\_Old\_Ack\_recd$}
     \label{discard}
        \If {Discard Old Ack is received from all affected switches}
         \LineComment{The update is complete.}
         \label{c_complete}
        \EndIf 
  \EndIf
\EndWhile
\EndProcedure
\end{algorithmic}
\end{algorithm}
\begin{algorithm}
\caption{Algorithm at the switch}
\label{Switch}
\begin{algorithmic}[1]
\Procedure {Switch}{null} 
\While {1}
   \If {$event$ = $commit\_recd$} \label{sstate1}
      \LineComment{$R_1$ and $R_0$ are received in Commit}
      \LineComment{\textbf{Begin atomic actions}}
      \For {each rule $r$ of $R_0$} 
         \State{Modify $f_2$ and $f_{p2}$ of $r$ to check for $0$}
         \State{Set $flag$ of $r$ $= OLD$}
      \EndFor
      \For {each rule $r$ of $R_1$}
         \State{Install $r$} \Comment{$f_1$ and $f_{p1}$ check for $0$}
         \State{Set $flag$ of $r$$ = NEW$}
      \EndFor
      \LineComment{\textbf{End atomic actions}}
      \State {$T_i$ = current time}
      \State {Send ReadyToCommit($v$,$T_i$)} \label{end_sstate1}
   \ElsIf {$event$ = $commit\_OK\_recd$} \label{sstate2}
      \State {$T_{last}$ = value of time received in Commit OK}
      \LineComment{\textbf{Begin atomic actions}}
      \For {each rule $r$ of $R_0$}
         \State{Set $T$ associated with $r$$ = T_{last}$}
      \EndFor
      \For {each rule $r$ of $R_1$}
         \State{Set $T$ associated with $r$$= T_{last}$}
      \EndFor
      \LineComment{\textbf{End atomic actions}}
      \State{$T_i$ = current time}
      \State{Send AckCommitOK($v$,$T_i$)}
      \label{end_sstate2}
   \ElsIf {$event$ =$discard\_old\_recd$}
      \State {$T_{del}$ = value received in Discard Old}
      \State{$T_i$ = current time}
      \State{$M$ = maximum time taken for a packet to be removed from the
network}
      \State \Call {start\_timer}{$T_{del}+M-T_i, v$} \label{start_timer}
      \State {Send DiscardOldAck($v$)}
   \ElsIf {$event$ = $timer\_expiry$}
      \label{timer_expiry}
      \State {Delete $R_0$ belonging to $v$ from the switch}
      \For {each rule $r$ of $R_1$}
         \State{Modify $f_1$,$f_2$,$f_{p1}$,$f_{p2}$ of $r$ to check for $*$}
         \State{Set $flag = U$}
      \EndFor
      \label{end_timer_expiry}
      \LineComment{The update is complete. After $M$ units of time, another
update conflicting with $v$ may begin.}
      \label{s_complete}
    \EndIf
\EndWhile
\EndProcedure
\end{algorithmic}
\end{algorithm}

\textbf{Algorithm at the control plane:}
This section first specifies the algorithm for the control plane, at the
controller (Algorithm~\ref{Controller}) and at each $s_i \in S$
(Algorithm~\ref{Switch}) and later for the data plane, in
Algorithm~\ref{Rule-Actions}.  The message exchanges below are the same as
in \cite{sukapuram2015enhanced} and \cite{CCU}; the parameters in them and
actions upon receiving them have been modified to suit PPCU.
\begin{enumerate}
\item \label{step1}{The Controller receives an Update Request from the
application, with the list of affected switches $S$, $R_0$ and $R_1$, for
every switch $s \in S$. The actions of all rules are as in Algorithm
\ref{Rule-Actions}. An update identifier $v$ is associated with each
update. The controller sends a ``Commit'' message to every switch $s \in S$
(line \ref{commit_sent} of Algorithm~\ref{Controller}) with $v$, $R_0$ and
$R_1$ as its parameters.\footnote{It is assumed that the controller and the switch
maintain all parameters related to an update in a store and check whether
each message received is appropriate; we omit these details and the error
handling required for an all-or-nothing semantics to improve clarity.}} 

\item\label{step2}{A switch $s \in S$ receives the ``Commit'' message,
extracts $v$, $R_0$ and $R_1$.  The switch 1) changes the match part of the
installed $R_0$ rules to check if $f_2 =0$ and $f_{p2}=0$ 2) installs the $R_1$ rules and
3) sets $flag = OLD$ for the $R_0$ rules and $flag = NEW$ for the
$R_1$ rules. The above actions must be atomic, as indicated in the
algorithm. Now it sends ``Ready to Commit'' with the current time of the
switch $T_i$ as a parameter (lines \ref{sstate1} to
\ref{end_sstate1} in Algorithm~\ref{Switch}).}

\item{The controller, upon receiving ``Ready To Commit'', stores the current
time received. Let $T_{last}$ be the largest value of time received in ``Ready
To Commit'' (Algorithm~\ref{Controller}, line~\ref{tlm}). 
Now it sends ``Commit OK'' to all
switches $s \in S$ with $T_{last}$ (Algorithm~\ref{Controller},
line~\ref{commit_ok_sent}).}

\item{Upon receiving ``Commit OK'', the switch sets $T = T_{last}$
in $R_0$ and $R_1$, which is considered a single atomic action. It now sends the current time
$T_i$ in ``Ack Commit OK''. See lines~\ref{sstate2} to \ref{end_sstate2} in
Algorithm~\ref{Switch}.}

\item{The controller receives ``Ack Commit OK'' from all the switches. Let
the largest value of time received in ``Ack Commit OK'' be $T_{del}$. It sends
``Discard Old'' to all the switches $s$ with $T_{del}$. }

\item{Let the time at which a switch $s$ receives ``Discard Old'' be $T_i$.
The switch sends ``Discard Old Ack'' to the controller.  The switch starts a
timer whose value is $T_{del}+ M - T_i$, where $M$ is the maximum lifetime
of a packet within the network and $T_{del}$ the time received in ``Discard
Old''.  (line~\ref{start_timer} in Algorithm~\ref{Switch}).  When the timer
expires, all the packets that were switched using the rules $R_0$ are no
longer in the network.  Therefore the switch deletes $R_0$. 
It sets $flag = U$ for every $R_1$ rule and modifies its $f_1$, $f_2$,
$f_{p1}$ and $f_{p2}$ to
check for $*$, as shown in 
lines~\ref{timer_expiry} to \ref{end_timer_expiry}, in Algorithm
~\ref{Switch}.  With this, the update at the switch is complete.}

\item{After the controller receives ``Discard Old Ack'' messages
from all the switches, the update is complete (line~\ref{c_complete} in
Algorithm~\ref{Controller}) at the controller.}
\end{enumerate}
Note: After $M$ units after timer expiry at the last affected
switch, the last of packets tagged $f_{p2}=1$ will exit the network. Now the
next update not disjoint with the current one may begin.

\textbf{Algorithm at the data plane - actions at switches:}
\begin{algorithm}
\caption{Rule actions}
\label{Rule-Actions}
\begin{algorithmic}[1]
\Procedure{Rule-Actions}{$action\_params$}
\LineComment{This is only the action part of any rule. 
$flag$ indicates if
the rule is new, old or unaffected. 
For a given value of $flag$, the match-field
values that the rule needs to match are below. $action\_params$ are passed from the rule.}
\If {$flag = NEW$} \label{new} \Comment{Match-field $f_1 = 0$ and $f_{p1}=0$}
\LineComment{An update is in progress and the rule is new}
   \If {$TS \geq T$ \textbf{OR} $f_{p2} = 1$} \Comment{Check time stamp of incoming packet} \label{alg-case1}
      \State{Set $f_{p2} = 1$} \Comment{A new packet} \label{alg-case3}
      \State{Execute actions with $action\_params$ }
\label{ex1}\Comment{$r_1$}
   \Else
      \LineComment{The packet will not match this rule again}
      \label{alg-case5}
      \State{Set $f_1 = 1$} 
      \LineComment{Use $recirculate$ if the change is to the egress
tables.} 
      \State{resubmit} \label{resub1} 
   \EndIf \label{end_new}
\ElsIf {$flag = OLD$}  \label{old} \Comment{Match-field $f_2 =0$ and
$f_{p2}=0$} \label{alg-case6}
\LineComment{An update is in progress and the rule is old}
   \If {$TS < T$ \textbf{OR} $f_{p1} = 1$} \label{alg-case4.1}
      \State{Set $f_{p1} = 1$} \Comment{An old packet} \label{alg-case4}
      \State{Execute actions with $action\_params$ }
\label{ex2}\Comment{$r_0$}
   \Else
      \LineComment{The packet will not match this rule again}
      \State{Set $f_2 = 1$} \label{alg-case7}
      \LineComment{Use $recirculate$ if the change is to the egress
tables.} 
      \State{resubmit} \label{resub2} 
   \EndIf \label{end_old}
\Else \Comment{Match-field $f_1=f_2=f_{p1}=f_{p2}=*$} \label{none}
   \If {$f_1 = 1$}
      \State{Set $f_{p1} = 1$} \label{alg-case5.1} \Comment{$R_1 - R_0 \neq \phi$}
   \ElsIf {$f_2 = 1$}
      \State{Set $f_{p2} = 1$} \label{alg-case7.1} \Comment{$R_0 \sim R_1 \neq \phi$}
   \EndIf
   \State{Execute actions with $action\_params$ } \label{ex3}\Comment{$r_u$}
\EndIf \label{end_none}
\EndProcedure
\end{algorithmic}
\end{algorithm}
Algorithm \ref{Rule-Actions} specifies the template for a compound action
associated with \textit{every} rule in a match-action table in a switch. 
Each rule has two metadata fields of $1$ bit each, $f_1$ and $f_2$,
associated with it, initialised to $0$ by default, indicating that the
packet is entering that table in that switch for the first time (If other
tables in the same switch has rule updates, a separate set of two bits need
to be used for \textit{each of those} tables. For ease of exposition, we
describe only one table being updated). Each rule has two registers,
$T$, initialised to $T_{max}$, which is $1$ less than the maximum value that $TS$ can
have, and $flag$, which decides if the rule is old ($OLD$), new ($NEW$) or
unaffected ($U$), initialised to $U$. Unaffected rules have $f_1$, $f_2$,
$f_{p1}$ and $f_{p2}$ set to $*$ in their match-fields and do not check for the
value of $TS$.  A new rule is always installed with a priority higher than
that of an old rule. In P4, rules with ternary matches have priorities
associated with them as such rules can have overlapping entries.

A rule may execute any of its own actions, which is what is
referred to in \ref{ex1}, \ref{ex2} and \ref{ex3} in Algorithm
\ref{Rule-Actions}. The parameters $action\_params$ that are passed to the
compound action are in turn passed to the actions associated
with the rule. For example, a new rule may require a packet to be forwarded
to port $5$, instead of port $6$. Then, the table entry for the new rule
would pass $5$ as a parameter from its rule, while the old rule would pass
$6$. 

Suppose $s_f$, the first affected switch of affected packet $p_1$ in Rules
Update $U_1$, receives ``Commit''. $TS < T_{max}$ of $p_1$ and let $R_1 -
R_0 =\phi$ for $s_f$. $p_1$ will match $r_1$ in $s_f$, get its $f_1$ set to $1$ and
get resubmitted (line \ref{resub1}). Now it will not match $r_1$ and instead
match $r_0$, get its $f_{p1}$ set to $1$ (line \ref{alg-case4}) and get
switched by $r_0$ in subsequent affected switches and $r_u$, if no $r_0$
exists. After $s_f$ receives ``Commit OK'' and sets its $T=T_{last}$, when a
packet $p_2$ whose $T > T_{last}$ arrives at $s_f$, it gets switched by the
new rule, gets its $f_{p2}$ set to $1$ (line \ref{alg-case3}) and gets
switched by $r_1$ rules in all the subsequent affected switches and $r_u$ if
no $r_1$ exists. Suppose in Rules Update $U_2$, $s_f$ has $R_1 - R_0 \neq
\phi$ and receives ``Commit''.  Now a packet $p_3$ entering it with $TS <
T_{max}$ gets its $f_1$ set to $1$, and then resubmitted(line \ref{resub1}).
The resubmitted packet matches another rule $r_u$ and gets its $f_{p1}$ set
to $1$ (line \ref{alg-case5.1}), making the packet use $r_0$ rules in
subsequent switches and $r_u$, where no $r_0$ exists. After $s_f$ receives
``Commit OK'', a packet $p_4$ whose $T > T_{last}$ matches $r_1$, gets its
$f_{p2}$ set to $1$ and gets switched in subsequent switches by $r_1$ or
$r_u$, if $r_1$ does not exist. The case for an update where $s_f$ has $R_0
\sim R_1 \neq \phi$ is similar.

\textbf{Specific scenarios:} We shall use a list (inexhaustive) of scenarios
to illustrate that each step in Algorithm \ref{Rule-Actions} is necessary:
Consider two switches $s_f \in S$ and $s_j \in S$, in Figures \ref{race1}
and \ref{r1-r0}, far apart from each other in the network. In all the cases,
the lower priority rule forwards a packet to a port while a higher priority
rule forwards a packet to the same port and increments a field $F$ in the
packet.  The value of $TS$ and the rest of the packet header are shown in
boxes with dotted lines, before and after crossing $s_f$.  The numbers $1$,
$2$ and $3$ against the boxes indicate the sequence of occurrence of events.

\subsubsection{Case 1, new rules must not be immediately applied}
After receiving ``Commit'', if any $s_i$ installs $r_1$ earlier than others, as packets
will immediately begin to match the new rules, PPC will be violated.
Therefore, each rule $r_1 \in R_1$ must check if $TS \geq T$ where $T
= T_{max}$ before it executes the actions associated with $r_1$ (line
\ref{alg-case1}). 

\subsubsection{Case 2, when to switch to the new rules} $T$ must not be
$T_{max}$ for packets to ever start using $r_1$. That time must be after all
the affected switches have installed the new rules and is now ready to use
them, which is $T_{last}$. Therefore after receiving ``Commit OK'',
each affected switch will set $T = T_{last}$ by using the run time API and
start using the new rules if $TS \geq T$ of an incoming packet.  However,
this is insufficient to guarantee PPC.

\subsubsection{Case 3, $s_f$ receives ``Commit OK'' first} 
\begin{figure}[t]
 \caption{Race conditions - $R_1 - R_0$ and $R_0 \sim R_1$ are $\phi$}
 \centering 
 \includegraphics[scale=0.4, trim=5 400 0 0]{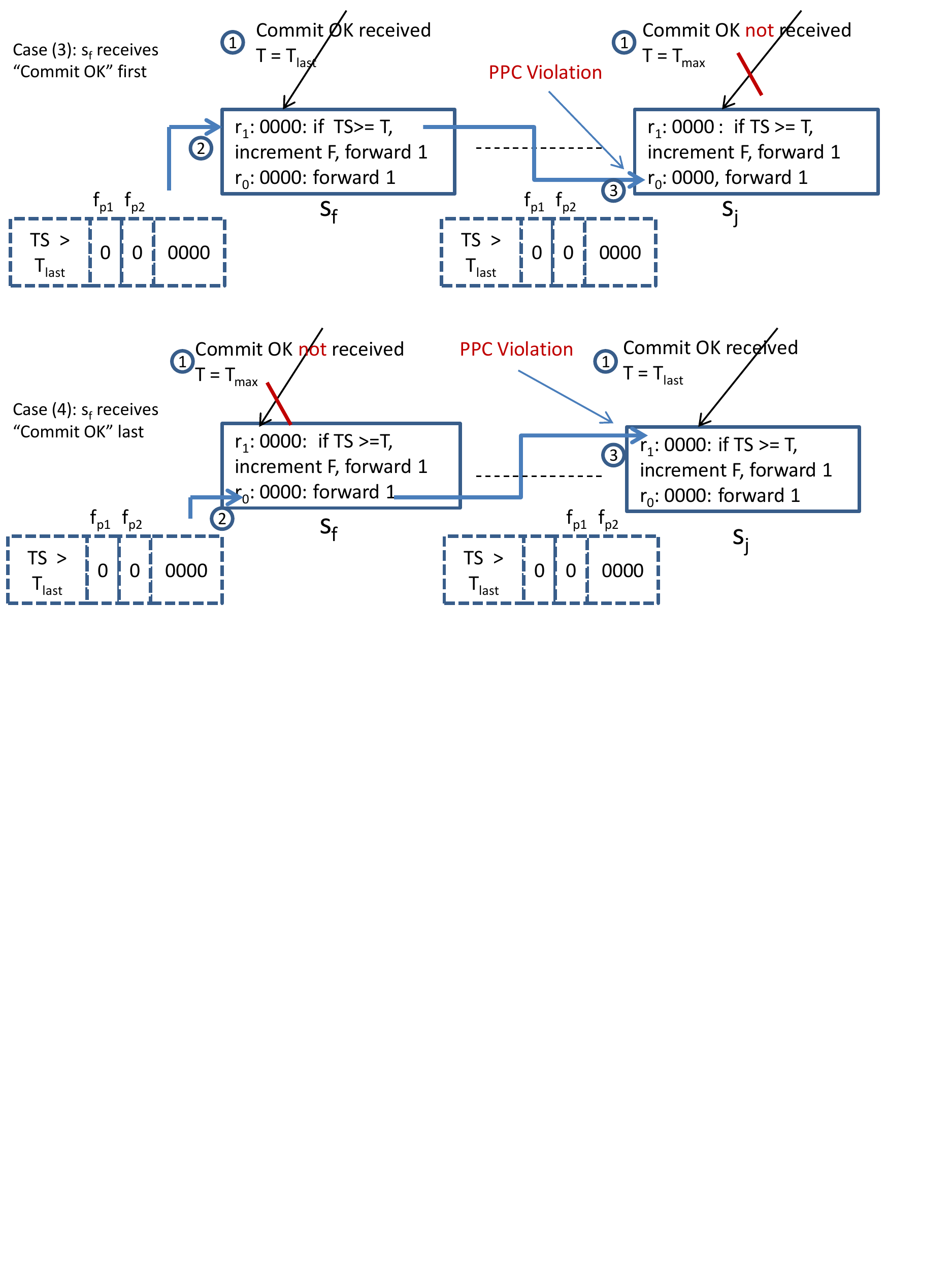} 
 \label{race1} 
\end{figure}

See 
Figure \ref{race1}. Suppose a packet $p$ whose $TS > T_{last}$ arrives at
the first affected switch $s_f$. Since $s_f$ has received ``Commit OK'', it
switches the packet using $r_1$ rules. Now the next affected switch $s_j$
has not received ``Commit OK'' yet. Therefore $r_1$ in $s_j$ checks if $TS
\geq T_{max}$, and since that is not so, $p$ gets switched using $r_0$
rules, thus violating PPC. To solve this problem, $r_1$ of $s_f$ must set
$f_{p2} = 1$ of all incoming packets that match $r_1$ and whose $TS \geq T$.
$r_1$ must check if $f_{p2}=1$ in the action (lines \ref{alg-case1} and
\ref{alg-case3}).

\subsubsection{Case 4, $s_f$ receives ``Commit OK'' last}
\label{case4}
To solve this problem, illustrated in Figure \ref{race1}, $r_0$ must set
$f_{p1} = 1$ and check for $f_{p1} = 1$ in the action (lines
\ref{alg-case4.1}, \ref{alg-case4}). $r_1$ must check for $f_{p1} =0$ in
its match field , so that $r_1$ rules in subsequent affected switches do not
match this packet and therefore always match $r_0$.

\subsubsection{Case 5, $R_1 - R_0 \neq \phi$ for $s_f$ and $s_j$}
\label{case5} 
To solve this issue, illustrated in Figure \ref{r1-r0}, the switch must
set $f_{p1}=1$ for packets whose $TS < T$ and matching $R_1 - R_0$ (lines
\ref{alg-case5} to \ref{resub1}, line \ref{alg-case5.1}). For this, the
switch first sets $f_1 =1$ and then resubmits the packet. Since $r_1$ checks
for $f_1=0$, the packet does not match $r_1$.  No $r_0$ exists for this
packet. Therefore the packet matches $r_u$, as intended. In $r_u$, if
$f_1=1$, $f_{p1}$ is set to $1$.

\subsubsection{Case 6, $R_0 \sim R_1 \neq \phi$ for $s_j$}
\label{case6}

\begin{figure}[t]
 \caption{Race conditions - One or more of $R_1 - R_0$ or $R_0 \sim R_1$ is $\phi$}
 \centering 
 \includegraphics[scale=0.4, trim=0 176 0 0]{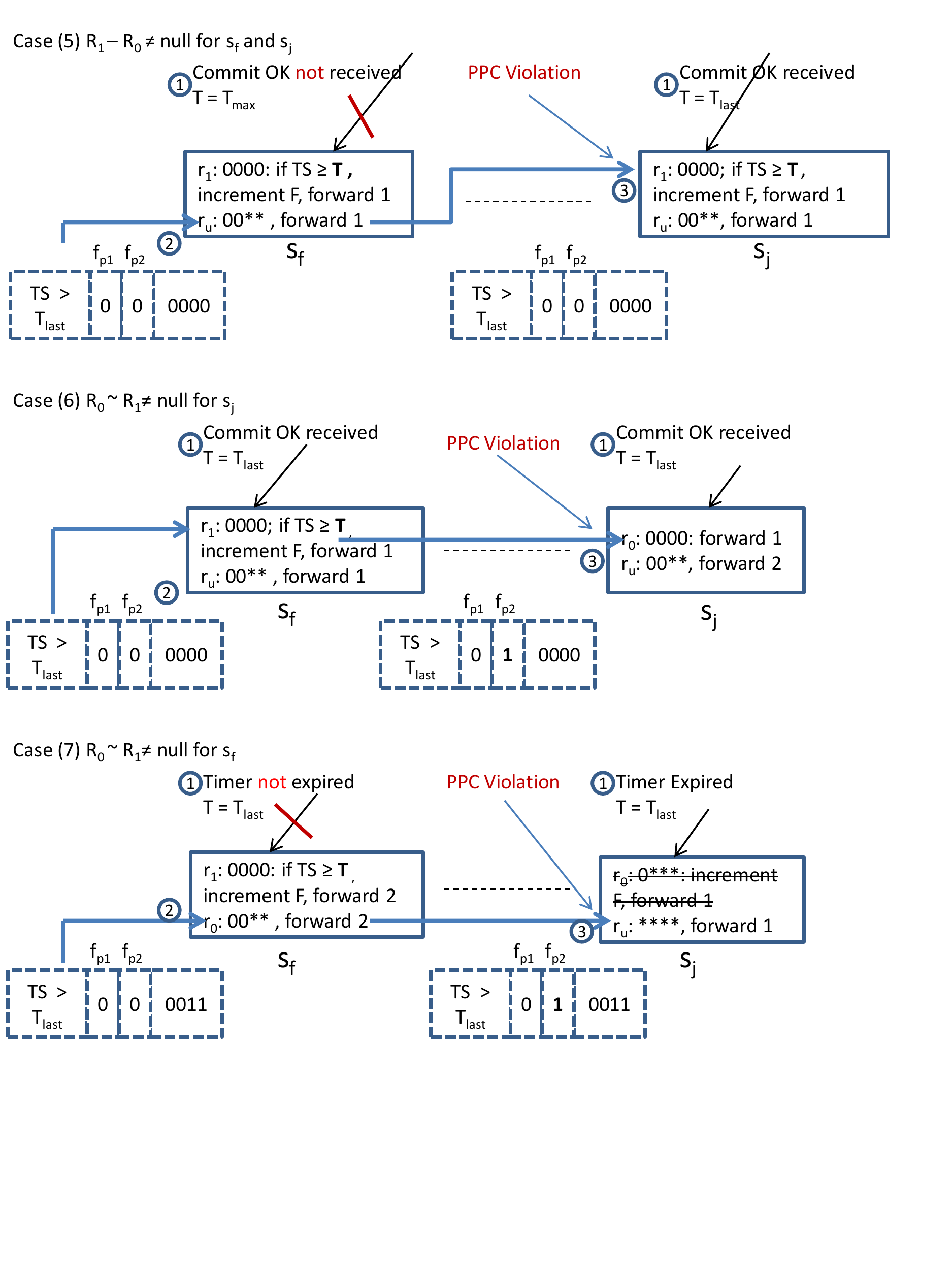} 
 \label{r1-r0} 
\end{figure}

To prevent a PPC violation, in  Figure \ref{r1-r0}, all old rules $r_0 \in
R_0$ must check if $f_{p2}=0$ in its match field (line \ref{alg-case6}).
Thus if a packet whose $TS \ge T_{last}$ arrives, it will not use any rule
$r_0 \in R_0$, instead it will use the next matching rule.

\subsubsection{Case 7, $R_0 \sim R_1 \neq \phi$ for $s_f$} \label{case7} See
Figure \ref{r1-r0}. $r_0$ gets deleted from $s_j$ after its
timer expires (line \ref{timer_expiry} in Algorithm \ref{Switch}).  The
solution for the issue here is to  set $f_{p2}=1$ of packets that match
$r_0$ as well, if their $TS \ge T_{last}$ (lines \ref{alg-case7} to
\ref{resub2} and line \ref{alg-case7.1}). Packets with $f_{p2}=1$ will use
$r_1$ rules if they match and unaffected rules if they do not, as unaffected
rules do not check for $f_{p1}$ or $f_{p2}$. Similar to the explanation in
section \ref{case5}, such a packet needs to be resubmitted after setting
$f_2 = 1$ and $r_0$ must check for $f_2 = 0$. Subsequently, in $r_u$, the
packet has its $f_{p2}$ set to $1$.  Since metadata instances are always
intialised to $0$ \cite{P4Spec}, unaffected rules need not set $f_1$ and
$f_2$ to 0.

\label{resubtime}
\textbf{On resubmitting packets:} It must be noted that packets need to be
resubmitted 1) only at $s_f$ and 2) only after receiving ``Commit OK'' and
until the timer expires, if $s_f$ has $R_0 \sim R_1 \neq \phi$ for an
update. In all other cases, packets need to be resubmitted 1) only at $s_f$
and 2) only after receiving ``Commit'' and until the switch receives
``Commit OK''. The action $recirculate$ must be used instead of $resubmit$,
if the change is to a table in the egress pipeline. 

\textbf{Impact of multiple tables:}
The algorithm requires no changes to support updates to more than one
ingress or egress table in a switch. The update to each table must be
atomic with respect to the update to another table within the same
switch. For example, when a switch receives ``Commit OK'', it must set $T =
T_{last}$ to the affected rules in all the tables, for that Rules Update,
atomically. How a switch implementation addresses this is outside the scope
of this paper. \cite{han2015blueswitch} is a line of work in this direction.
\section{Concurrent Updates}
Each disjoint Rules Update requires a unique update identifier $v$ for the
duration of the update, to track the update states at the affected switches
and the controller. The number of disjoint Rule Updates that can be
simultaneously executed is limited only by the size of $v$. $v$ is exchanged
only between the controller and the switches and hence is not dependent on
the size of a field in any data packet. Therefore as many disjoint updates as the
size of the update identifier or the processing power of switches would
allow can be executed concurrently.
\section{Using time stamps}
\textbf{Feasibility of adding $TS$ at the ingress:} We assume that all
switches have their clocks synchronized and the maximum time drift $\gamma$
of switches from each other is known. If the network supports Precision Time
Protocol, $\gamma = 1 \mu sec$ \cite{mizrahi2016software}.  Intel FM6000, a
\textit{programmable} SDN capable switch, supports PTP, its $\gamma<1\mu
sec$ and the time stamp is accessible in software \cite{fm6000}.  The size
of the register used to store a packet timestamp in FM6000 is $31$ bits. P4
supports a feature called ``intrinsic metadata'', that has target specific semantics
and that may be used to access the packet timestamp. Using the $add\_header$ and
$remove\_header$ actions and intrinsic metadata, the $TS$ field may be added
at the ingress and removed at the egress for every packet, for targets that
support protocols such as PTP.

\textbf{Asynchronous time at each switch:} Since a single rule update in a
TCAM is of the order of milliseconds, a time stamp granularity of
milliseconds is sufficient for Rules Updates.  Let us assume that the
timestamp value is in milliseconds and that an ingress $s_i$ is
\textit{faster} than an affected internal switch (at the most by $1 \mu
sec$).  Let $t_1$ be the time stamp of a packet $p$ in the network that is
stamped by $s_i$, even before the Rules Update begins.  After the Rules
Update begins, let the (temporally) last affected switch send its current
time stamp $T_{last}$ in ``Ready To Commit''. Let $t_1 > T_{last}$, since
the ingress is faster. Now a switch $s_j \in S$, which is not an $s_f$, will
switch $p$ with the new rules, violating PPC. To prevent this, each switch
may set $ T= \left \lceil{T_{last}+\gamma + 1}\right \rceil$, instead of $T =
T_{last}$. If an ingress clock is slow, there will be no PPC violations.
Thus PPCU tolerates known inaccuracies in time synchronisation.  To reduce
packet transmission times, the size of $TS$ may be reduced, thus reducing
the granularity. $s_f$ will need to wait longer to receive packets whose $TS
\geq T_{last}$, thus lengthening the Rules Update time. Since the size of
$TS$ is programmable in the field, the operator may be choose it according
to the nature of the network.
\section{Proof}
\label{proof}
We need to prove that the algorithms \ref{Controller}, \ref{Switch} and
\ref{Rule-Actions} together provide PPC updates. $p(m,f_{p1},f_{p2})$
denotes an affected packet with match-field $m$ and fields $f_{p1}$ and
$f_{p2}$.  $r(m)$ denotes a rule whose match-field is $m$ (in both, $m$
excludes $TS$, $f_{p1}$ and $f_{p2}$).  By definition of $S$, for a packet
$p(m,f_{p1},f_{p2})$, any $s \in S$ has either a rule $r_0(m)$ or $r_1(m)$
or both.

Let us assume that the individual algorithms \ref{Controller}, \ref{Switch}
and \ref{Rule-Actions} are correct. Let us assume that switches are
synchronized, to make descriptions easier. In the description below, when we
refer to a packet, we mean a packet \textit{affected by the update}. The
update duration is split into three intervals: before $s_f$
receiving ``Commit'', from $s_f$ receiving ``Commit'' to the timer expiring,
and after the timer expiry. 

\textbf{Property 1: After the Rules Update begins and before $s_f$ receives
``Commit'', PPC is preserved:} \label{L1} $s_f$ (and all $s_j$ that have not
received ``Commit'') switches all packets using a rule that will be marked
$r_0(m)$ on receiving ``Commit'' or $r_u(m)$ (if no rule is to be deleted on
that switch). The first $s_j$ that receives ``Commit'' switches using
$r_0(m)$ or $r_u(m)$ (the latter if the switch does not have $r_0(m)$ but
has only $r_1(m)$) and sets $f_{p1}=1$ for all received packets.  Subsequent
$s_j$s will not use $r_1(m)$ as $f_{p1}=1$ and use $r_0(m)$ if it exists (or
$r_u(m)$ if no $r_0(m)$ exists) .  Thus packets follow $r_0(m)$ if it exists
or $r_u(m)$ if $r_0(m)$ does not exist, preserving PPC.

\textbf{Property 2: Between $s_f$ receiving ``Commit'' at $s_f$ and its timer
expires PPC is preserved:}
\label{L2}
Property 2.1: As per Algorithm \ref{Rule-Actions}, if $s_f$ has received
``Commit'', $p(m,f_{p1},f_{p2})$ will exit $s_f$ as either $p(m,0,1)$ or
$p(m,1,0)$, until its timer expires.

Property 2.2: $s_j$ will receive any of the types of packets $1$) $p(m,0,0)$
with its $TS < T_{last}$ or $2$) $p(m,1,0)$ or $3$) $p(m,0,1)$ and no other type
of packet, until its timer expires.  Proof: Packets of type $1$ may reach $s_j$
as it may have crossed $s_f$ before $s_f$ received ``Commit'' and $TS$ will
always be less than $T_{last}$ of such packets, as $T_{last}$ is the time at
which the last switch has received ``Commit''. If a packet of type other
than these three reaches $s_j$, it means it has not crossed $s_f$, by
Property 2.1. In that case, the only possibility is that $s_j$ is the first
switch, which is not true by defintion.

By Algorithm \ref{Rule-Actions}, packets of types $1$ and $2$
referred to in Property 2.2 can use only $r_0$ rules (or $r_u$ if $r_0$
does not exist) on all switches $s_j$.  By the same algorithm, packets of
type $3$ match only $r_1$ rules (or $r_u$ if $r_1$ does not exist) on all
switches $s_j$.  In $s \notin S$, packets match $r_u$. $r_u$ can change the
value of $f_{p1}$ or $f_{p2}$ of  a packet only if $f_1$ or $f_2$ for that
packet is set, which occurs only on an affected switch. Therefore, PPC is
preserved.

\textbf{Property 3: When the timer expires at any switch $s_i \in S$, all packets
that were switched using $r_0$ at least at one $s \in S$ have left the
network:}
\label{L3}
Let us consider two cases: 1) All packets whose $TS < T_{last}$ get switched
by $s_f$ using $r_0$ rules, by Algorithm \ref{Rule-Actions}.  Such  packets
exit the network by time $T_{last} + M$. However, $T_{del} > T_{last}$. 2)
All packets whose $TS \geq T_{last}$ that arrive at $s_f$ get switched using
$r_0$, until that $s_f$ receives its ``Commit OK'', by Algorithm
\ref{Rule-Actions}. At time $T_{del}$, the last of $s_f$s have received
``Commit OK'' and have started switching packets using $r_1$. Hence at time
$T_{del} + M$, all the packets (whether due to case 1 or 2) that were
switched using $r_0$ have left the network.  The current time at a switch is
$T_i$ and the switch receives an instruction to start a timer that expires
at $T_{del} + M$. The elapsed time, that is, $T_i - T_{del}$ must be
subtracted from $M$ , giving the timer a value of $M-T_i + T_{del}$.
Therefore when this timer expires, all the old packets have left the
network.

\textbf{Property 4: After timer expiry at $s_f$:}
\label{L4}
All packets in the network have their $TS \geq T_{last}$ after timer expiry
at $s_f$, as all packets whose $TS < T_{last}$ have left the network by the
timer expiry at any switch, by Property 3. Therefore all packets leaving
$s_f$ are of the form $p(m,0,1)$. Therefore all switches $s_j \neq s_f$ will
use $r_1$ if it exists (or $r_u$ if $r_1$ does not exist), if the timer has
not expired on $s_j$. If the timer has expired on $s_j$, it will only cause
it to not check for $f_{p1}$, $f_{p2}$, $f_1$ or $f_2$ , which makes no
difference to matching the rule that the packet matches. Since packets have
only $r_1$ to match (or $r_u$ where it does not exist), PPC is preserved.

Thus due to Properties 1 to 4, PPC is preserved during and after the update.

\section{Analysis of the Algorithm}
The symbols used in the analysis are: $\delta$: the propagation time between
the controller and a switch, $t_{i}$: the time taken to insert rules in a
switch TCAM, $t_{d}$: the time taken to delete rules from the TCAM, $t_{m}$:
the time taken to modify rules in the TCAM, $t_{v}$: the time taken to
modify registers associated with a rule, $t_s$: the time for
which a switch waits after it receives ``Discard Old'' and before it deletes
rules, $n_o$: the number of old rules that need to be removed, $n_n$: the
number of new rules that need to be added, $n$: the maximum number of rules
in a switch, $k_a$: the number of affected switches, $k_i$: the number of
ingresses, $k_t$: the total number of switches, $R_1$: The
time between the switch receiving ``Commit'' and sending ``Ready To
Commit'', $R_2$: The time between the switch receiving ``Commit OK'' and
sending ``Ack Commit OK'' and $R3$: The time between the switch receiving
``Discard Old'' and the switch performing its functions after timer expiry.
It is assumed that the value of $\delta$ is uniform for all switches and all
rounds, the values of time are the worst for that round, the number of
rules, the highest for that round and that the processing time at the
controller is negligible. Since unaffected rules have ternary matches for
$f_{p1}$ and $f_{p2}$, we assume that all the match fields are stored in
TCAM and the corresponding actions in SRAM \cite{bosshart2013forwarding}
\cite{jose2015compiling},
for PPCU, and for other algorithms.

\begin{table}[!t]
\caption{Comparison with E2PU and CCU}
\label{table:analysis}
\begin{tabular}{|p{1.75cm}|p{1.75cm}|p{1.75cm}|p{1.75cm}|} \hline
\textbf{Parameter}&\textbf{PPCU}&\textbf{E2PU\cite{sukapuram2015enhanced}}&\textbf{\cite{luo2015consistency}}\\
\hline
\hline
Message complexity & $6k_a$ & $4k_t$ & $4k_a + 4k_{i}$\\
\hline
FP & 1 & $k_a/k_t$ & $k_a / (k_a+k_i)$\\
\hline
Round 1 (R1) &$n_o(t_{m} + t_v) + n_n(t_{i} + t_{v})$ & $(n - n_o + n_n) t_i$ & $(n_o* t_m)+ (n_n * t_i)$\\
\hline
Round 2 (R2) & $(n_o + n_n)* t_v$ & $(n - n_o)t_m + n_n*t_i$ & $n * t_m$\\
\hline
Round 3 (R3)& $ t_s+ n_n * (t_m + t_v) + n_o * t_d$ & $ t_s +
n*t_d$ & $t_s + n_n*t_m+ n_o *t_d$\\
\hline
Round 4 & Not applicable & Not applicable & $n*t_m$\\
\hline
Propagation Time P & $6\delta$ & $6\delta$  & $8\delta$\\
\hline
Time Complexity & $P+R1+R2+R3$ & $P+R1+R2+R3$ & $P+R1+R2+R3+R4$\\
\hline
Concurrency & Unlimited & $0$ & Number of bits in version
field\\
\hline
\end{tabular}
\end{table}

We add to and evaluate using the parameters of interest identified for a PPC
in \cite{sukapuram2015enhanced}: 1) \textbf{Overlap}: Duration for which the
old and new rules exist at each type of switch 2) \textbf{Transition time}:
Duration within which new rules become usable from the beginning of the
update at the controller 3) \textbf{Message complexity:} the number of
messages required to complete the protocol 4) \textbf{Time complexity:} the
total update time\footnote{Excludes the time for current packets to be
removed from the network at the end of the update, where applicable}. 5)
\textbf{Footprint Proportionality} 6) \textbf{Concurrency}: Number of
disjoint concurrent updates.

The purpose of the analysis is to understand what the above depend upon and
to compare with a single update in E2PU (using rule updates in TCAMs)
\cite{sukapuram2015enhanced}, which updates switches using 2PU with 3 rounds
of message exchanges, and with the algorithm in \cite{luo2015consistency}.
We assume that \cite{luo2015consistency} uses acknowledgements for each
message sent. Since all three have similar rounds, it is meaningful to
compare the time taken by each round as given in table \ref{table:analysis}.
The Overlap is $4\delta+R1+R2+R3$ and Transition time is $3\delta+R1+R2$ for
all the algorithms under consideration.

Since the values associated with the action part of a rule ($T$ and $flag$)
are stored in SRAM, we assume that update times of these values (in $ns$)
will be negligible compared to TCAM update times and ignore the terms that
involve $t_v$. We find in table \ref{table:analysis} that 1) PPCU has lower
message complexity, assuming $k_i > k_a/2$ and better FP 2) PPCU and
\cite{luo2015consistency} have comparable times for Rounds 1 and 3.  For
Round 2, PPCU fares better. Therefore, PPCU fares better for Overlap,
Transition time and Time complexity. 3) PPCU has better concurrency. From
\ref{resubtime} it may be inferred that for PPCU, packets need to be
resubmitted either for a duration of $R2+2\delta+R3$ or $R1+2\delta+R2$ at
only $s_f$. Since $resubmit$ is an action supported by line rate switches,
we assume that the delay due to a resubmission at $s_f$ for this time frame
during an update is tolerable.

\textbf{Feasibility of implementation at line rate:} Adding and removing
headers such as $TS$, $f_{p1}$ and $f_{p2}$ are feasible at line rate; so
are setting and checking metadata, such as $f_1$ and $f_2$ and header fields
$f_{p1}$ and $f_{p2}$ in actions, as per table $1$ in RMT
\cite{bosshart2013forwarding}. While compiling the action part, Domino
\cite{packettx} checks if operations on stateful variables in actions can
run at line rate by mapping those operations to its instruction set - PPCU
requires only reading the state variables $flag$ and $T$ and this can be
achieved using the ``Read/Write atom'' (name of instruction) in Domino. This
demonstrates the feasibility of PPCU running at line rate.
\section{Conclusions}
PPCU is able to provide non-conflicting per-packet consistent
updates with an all-or-nothing semantics, with updates confined
to the affected switches and rules, with no restrictions on the scenarios
supported and no limits to concurrency. The algorithm fares better than
other algorithms that preserve PPC, in a theoretical evaluation of
significant parameters. PPCU uses only existing programming features for the
data plane, are implementable with the machine instruction sets
supported and therefore must work at line rate. Available programmable
switches support timing protocols such as PTP and time stamping packets is
possible at line rate, as required by the algorithm. It accomodates time
asynchrony, as long as the maximum time drift of switches from each other is
known, which is the case with PTP.

\bibliographystyle{IEEEtran}
\bibliography{IEEEabrv,newlit}
%



\end{document}